\documentclass[pra,twocolumn,showpacs,amssymb]{revtex4}
\usepackage{amssymb}
\usepackage{latexsym}
\usepackage{graphicx}
\usepackage{enumerate}
\usepackage{amsmath}
\usepackage{dcolumn}
\usepackage{bm}

\def\ba{\begin{eqnarray}}
\def\ea{\end{eqnarray}}

\def\be{\begin{equation}}
\def\ee{\end{equation}}

\begin{document}

\author{Graciana Puentes}
\affiliation{Departamento de Fisica, Facultad de Ciencias Exactas y
Naturales, Pabell\'{o}n 1, Ciudad Universitaria, 1428 Buenos Aires,
Argentina}

\title{Quantum Walk Topology and Spontaneous Parametric Down Conversion}

\begin{abstract}

In a recent detailed research program we  proposed to study the complex physics of topological phases by an all optical implementation of a discrete-time quantum walk. The main novel ingredient proposed for this study is the use of non-linear parametric amplifiers in the  network which could in turn be used to emulate intra-atomic interactions and thus analyze many-body effects in topological phases even when using light as the quantum walker. In this paper, and as a first step towards the implementation of our scheme, we analize the interplay between quantum walk lattice topology and spatial correlations of bi-photons produced by spontaneous parametric down-conversion. We also describe different detection methods suitable for our proposed experimental scheme. 


\end{abstract}

\maketitle
\section{INTRODUCTION}


\noindent Phase transitions play a fundamental  role in physics. From  melting ice  to the creation of mass in the Universe, phase transitions are at the center of most dynamical processes which involve an abrupt change in the properties of a system.~Phase transitions are usually driven by some form of fluctuation. While classical phase transitions are typically driven by thermal noise, quantum phase transitions are triggered by  quantum fluctuations. Quantum phase transitions have been extensively studied in a large number of fields ranging from cosmology  to condensed matter and have received much attention in the field of ultra-cold atoms since the observation of Bose-Einstein condensation \cite{BEC}, and the subsequent experimental realization of Superfluid-Mott Insulator phase transition in optical lattices \cite{Greiner}. A common feature of quantum phase transitions is that they involve some form of spontaneous symmetry breaking, such that the ground state of the system  has less overall symmetry than the Hamiltonian and  can be described by a \emph{local} order parameter. \\

\noindent A rather distinctive class of quantum phases is present in systems characterized by a Hilbert space which is split into different topological sectors, the so called topological phases.  Topological phases have received much attention after the discovery of the quantum Hall effect \cite{Thouless} and the interest increased following the prediction \cite{Kane} and experimental realization \cite{Koening} of a new class of material called topological insulators. Topological insulators are band insulators with particular symmetry properties arising from spin-orbit interactions which are predicted to exhibit surface edge states which should reflect the non-trivial topological properties of the band structure, and which should be topologically protected by time reversal symmetry. Unlike most familiar phases of matter which break different kinds of symmetries, topological phases are not characterized by a broken symmetry, they have degenerate ground states  which present more symmetry than the underlying Hamiltonian, and can not be described by a \emph{local} order parameter. Rather, these partially unexplored type of phases are described by topological invariants, such as the Chern number which is intimately related to the adiabatic Berry phase, and are predicted to convey a variety of exotic phenomena, such as fractional charges and magnetic monopoles \cite{Qi}. It has recently been theoretically demonstrated that it is possible to simulate a large ``Zoo" of topological phases by means of discrete-time quantum walks (DTQWs) of a single particle hopping between adjacent sites of an optical lattice, through a sequence of unitary operations \cite{Kitagawa,KitagawaNatComm}.\\

In this paper, we propose a detailed research program for the study for non-linear effects in photonic quantum walks and their interplay with topological phenomena \cite{PuentesJPB2012}. This paper is based on an original proposal written in the year 2010 by G. Puentes \cite{PuentesArxiv, PuentesArxiv2}. More specifically, we analyze the interplay between a non-trivial topology determined by a linear quantum walk Hamiltonian ($H_{\mathrm{QW}}$), on the phase-matching condition characterizing bi-photons produced by the non-linear process of spontaneous parametric down conversion (SPDC), characterized by a non-linear Hamiltonian ($H_{\mathrm{SPDC}}$). By considering both a linear and non-linear contributions in the overall bi-photon Hamiltonian, we analyze the coupling efficiency and emission probability in different topological scenarios.  \\

\noindent Random walks have been used to model a variety of dynamical physical processes containing some form of stochasticity, including  phenomena such as Brownian motion and the transition from binomial to Gaussian distribution in the limit of large statistics. The quantum walk (QW) is the quantum analogue of the random walk, where the classical walker is replaced by a quantum particle, such as a photon or an electron, and the stochastic evolution is replaced by a unitary process. The stochastic ingredient is added by introducing some internal degrees of freedom  which can be stochastically flipped during the evolution, which is usually referred to as a  quantum coin. One of the main ingredients of quantum walks is that the different paths of the quantum walker can interfere, and therefore present a complicated (non Gaussian) probability distribution for the final position of the walker after a number of  discrete steps. In recent years, quantum walks have have been successfully implemented to simulate a number of quantum phenomena such as photosyntesis \cite{Mohseni}, quantum diffusion \cite{Godoy}, vortex transport \cite{Rudner} and electrical brake-down \cite{Oka}, and they have provided a robust platform for the simulation of quantum algorithms and maps \cite{Paz}. QWs have been experimentally implemented in the context of NMR \cite{Ryan}, cavity QED \cite{Agarwal}, trapped ions \cite{ions}, cold atoms \cite{Karski} as well as  optics, both in the spatial \cite{Do} and frequency domain \cite{Bouwmeester}. In recent years, quantum walks with single and correlated photons have been successfully introduced using wave-guides \cite{Peruzzo} and bulk optics \cite{photons2}, and time-domain implementations \cite{photons}.\\

\noindent It is relevant to point out that any implementation of a quantum walk so far \cite{Godoy, Rudner, Oka, Paz, Agarwal, Karski, Bouwmeester, Peruzzo, photons2,photons, Silberhorn1D,Silberhorn} has introduced passive linear elements \emph{only} for the composing elements of the random network. 
A full class of topological insulators can be realized in a system of non-interacting particles, with a binary (psuedo) spin space for (bosons) fermions, via a random walk of discrete time unitary steps as described in Ref \cite{Kitagawa}. The particular type of phase is determined by the size of the system (1D or 2D) and by the underlying symmetries characterizing the Hamiltonian, such as particle-hole symmetry (PHS), time-reversal symmetry (TRS), or chiral symmetry (CS). The 1D discrete time quantum walk (DTQW) can be specified by a series of unitary spin dependent translations $T$ and rotations $R(\theta)$, where $\theta$ specifies the rotation angle. Thus, the quantum evolution is determined by applying a series of unitary operations or steps:

\begin{equation}
\label{eq:1}
U(\theta)=TR(\theta).
\end{equation}

\noindent The generator of the unitary evolution operator (or map) in Eq. [1] is the time-independent Hamiltonian $H(\theta)$, for which the discrete time evolution operator  $U(\theta)$ can be defined as:

\begin{equation}
U(\theta)=e ^{-iH(\theta)\delta t},
\end{equation}

\noindent where we have chosen $\hbar=1$, and the finite time evolution after $N$ steps is given by  $U^{N}=e^{-i H(\theta)N\delta t}$. \\

\noindent The Hamiltonian $H(\theta)$ determined by the translation and rotation steps $T$ and $R(\theta)$, posses particle hole symmetry (PHS)  for some operator $P$ (i.e. $PHP^{-1}=-H$) and it also contains chiral symmetry (CS). The presence of PHS and CS guaranties time reversal symmetry (TRS). The presence of TRS and PHS imply that the system belongs to a topological class contained in the Su-Schrieffer-Heeger (SSH) model \cite{Su} and can thus be employed to simulate a class of SSH topological phase.~An extension to 2D topological insulator can be obtained by extending the lattice of sites to 2D. Different geometries such as square lattice or triangular lattice are described in Ref \cite{Kitagawa}. In this work we propose to study the dynamical evolution given a general overall Hamiltonian of the form:

\begin{equation}
H=H_{\mathrm{QW}}+H_{\mathrm{SPDC}},
\end{equation} 

where the first term is the linear contribution given by the non-trivial topology of the quantum walk lattice, and the second term is the non-linear contribution due to the spontaneous parametric down conversion in non-linear media along the lattice. 

\section{Split-Step Quantum Walk Hamiltonian ($H_{\mathrm{QW}}$)}

The basic step in the standard DTQW is given by a unitary evolution operator $U(\theta)=TR_{\vec{n}}(\theta)$, where $R_{\vec{n}}(\theta)$ is a rotation along an arbitrary direction $\vec{n}=(n_{x},n_{y},n_{z})$, given by: 
\begin{equation}
R_{\vec{n}}(\theta)=
\left( {\begin{array}{cc}
 \cos(\theta)-in_{z}\sin(\theta) & (in_{x}-n_{y})\sin(\theta)  \\
 (in_{x}+n_{y})\sin(\theta) & \cos(\theta) +in_{z}\sin(\theta)  \\
 \end{array} } \right),
\end{equation}

 in the Pauli basis \cite{Pauli}. In this basis, the y-rotation is defined by a coin operator of the form  
\begin{equation}
R_{y}(\theta)=
\left( {\begin{array}{cc}
 \cos(\theta) & -\sin(\theta)  \\
 \sin(\theta) & \cos(\theta)  \\
 \end{array} } \right)
\end{equation}
 This is  
followed by a spin- or polarization-dependent translation $T$ given by 

\begin{equation}
T=\sum_{x}|x+1\rangle\langle x | \otimes|H\rangle \langle H| +|x-1\rangle \langle x| \otimes |V\rangle \langle V|,
\end{equation}

 where $H=(1,0)^{T}$ and $V=(0,1)^{T}$. The evolution operator for a discrete-time step is equivalent to that generated by a Hamiltonian $H(\theta)$, such that $U(\theta)=e^{-iH(\theta)}$ ($\hbar=1$), with $H_{QW}(\theta)=\int_{-\pi}^{\pi} dk[E_{\theta}(k)\vec{n}(k).\vec{\sigma}] \otimes |k \rangle \langle k|$ and $\vec{\sigma}$ the Pauli matrices, which readily reveals the spin-orbit coupling mechanism in the system.~The quantum walk described by $U(\theta)$ has been realized experimentally in a number of systems \cite{photons,photons2,ions, coldatoms}, and has been shown to posses chiral symmetry, and display Dirac-like dispersion relation given by $\cos(E_{\theta}(k))=\cos(k)\cos(\theta)$.\\

Here we analize a DTQW protocol consisting of two consecutvie spin-dpendent translations $T$ and rotations $R$, such that the unitary step becomes $U(\theta_1,\theta_2)=TR(\theta_1)TR(\theta_2)$. The so-called ``split-step" quantum walk, has been shown to possess a non-trivial topological landscape characterized by topological sectors with different topological numebers, such as the winding number $W=0,1$. The dispersion relation for the split-step quantum walk results in \cite{Kitagawa}:

\begin{equation}
 \cos(E_{\theta,\phi}(k))=\cos(k)\cos(\theta_1)\cos(\theta_2)-\sin(\theta_1)\sin(\theta_2).
\end{equation}

~The 3D-norm for decomposing the quantum walk Hamiltonian of the system in terms of Pauli matrices $H_{\mathrm{QW}}=E(k)\vec{n} \cdot \vec{\sigma}$  becomes \cite{Kitagawa}: \\
\begin{equation}
\begin{array}{ccc}
n_{\theta_1,\theta_2}^{x}(k)&=&\frac{\sin(k)\sin(\theta_1)\cos(\theta_2)}{\sin(E_{\theta_1,\theta_2}(k))}\\
n_{\theta_1,\theta_2}^{y}(k)&=&\frac{\cos(k)\sin(\theta_1)\cos(\theta_2)+\sin(\theta_2)\cos(\theta_1)}{\sin(E_{\theta_1,\theta_2}(k))}\\
n_{\theta_1,\theta_2}^{z}(k)&=&\frac{-\sin(k)\cos(\theta_2)\cos(\theta_1)}{\sin(E_{\theta_1,\theta_2}(k))}.\\
\end{array}
 \end{equation}
\bigskip

Diagonalization of $H_{\mathrm{QW}}$ gives the lattice Bloch eigen-vectors, characterizing the quantum walk Hamiltonian, result in:

\begin{equation}
{\bf{u}_{\pm}}(k)=\frac{1}{\mathcal{N}}(1,\frac{n_{x}(k)+in_{y}(k)}{n_{z}(k)\pm\lambda(k)})^{T},
 \end{equation}

with $ \lambda^2=n_{x}^2+n_{y}^2+n_{z}^2$, and $\mathcal{N}$ a normalization factor. We note that the relation between the two components of ${\bf{u}_{\pm}}$ will eventually determine the phase-matching condition for down converted photons, and for this reason it is of relevance for our analysis. \\

For the particular case that $n_{z}(k)=0$, the eigen-vectors take the simple form:
\begin{equation}
{\bf{u}_{\pm}}(k)=\frac{1}{\sqrt{2}}(1,e^{-i\phi(k)})^{T},
 \end{equation}

with $\phi(k)=\mathrm{atan}(\frac{n_{y}}{n_{x}})$. For the split-step  quantum walk this relative phase results in:

\begin{equation}
 \phi(k)=\mathrm{atan}(\frac{\cos(k)\sin(\theta_1)\cos(\theta_2)+\sin(\theta_2)\cos(\theta_1)}{\sin(k)\sin(\theta_1)\cos(\theta_2)}).
\end{equation}

\subsection{Zak Phase}

We can gain further insight by calculating the Zak phase of this system, which is analogous to the Berry phase on the torus (i.e., the Brillouin zone). Consider the general Hamiltonian:
\begin{equation}
H\sim n_x \sigma_x+n_y \sigma_y+ n_z \sigma_z,
\end{equation}
Since the eigenvalues are the only quantities of interest for the present problem, the overall constants of this Hamiltonian can be safely ignored.
The explicit expression for this Hamiltonian is
\begin{equation}
H=
\left(
\begin{array}{cc}
  n_z \qquad    n_x-in_y\\
 n_x+i n_y \qquad   -n_z   
\end{array}
\right),
\ee
with eigenvalues
\begin{equation}
\lambda=\pm \sqrt{n_x^2+n_y^2+n_z^2}
\end{equation}
The normalized eigenvectors then result
\begin{equation}
|V_\pm>= 
\left(
\begin{array}{cc}
  \frac{n_x+i n_y}{\sqrt{2n_x^2+ 2n_y^2+2n_z^2\mp 2n_z\sqrt{n_x^2+n_y^2+n_z^2}}}    \\
  \frac{n_z\mp \sqrt{n_x^2+n_y^2+n_z^2}}{\sqrt{2n_x^2+ 2n_y^2+2n_z^2\mp 2n_z\sqrt{n_x^2+n_y^2+n_z^2}}}   
\end{array}
\right)
\end{equation}

Note that the scaling $n_i\to\lambda n_i$ does not affect the result, as should be. The overall Zak phase for the problem is:

\begin{equation}
Z=i\int (<V_+|\partial_k V_+>+ <V_-|\partial_k V_->) dk.
\end{equation}

For the split-step quantum walk, the Zak phase results in an analytic expresssion of the form \cite{PuentesArxiv2}:

\begin{equation}
Z=\phi(-\pi/2)-\phi(\pi/2)=\frac{\tan(\theta_2)}{\tan(\theta_1)}.
\end{equation}

\section{Spontaneous Parametric Down-Conversion (SPDC) Hamiltonian ($H_{\mathrm{SPDC}}$)}

We can decompose the $H_{\mathrm{SPDC}}$ in terms  of the Bloch eigen-vectors ${\bf{u}_{\pm}}(k)$ by defining Bloch waves of the form $\hat{A}_{p}(k)=\sum_{n} {\bf{\hat{a}_{n}}}{\bf{u}_{\pm}}(k)e^{ikn}$, with $\bf{\hat{a}_{n}}$=$(\hat{a}_{n,1},\hat{a}_{n,2},...,\hat{a}_{N,m})$ and $\hat{a}_{N,m}$  the anihilation operator of the $m^{th}$ sublattice. The SPDC Hamiltonian results in:

\begin{equation}
 H_{\mathrm{SPDC}}=\sum_{p_{s,i}} \int {dk_{s}}\int {dk_{i}}\Gamma_{p_{s,i}}(k_{s},k_{i})\hat{A^{\dagger}}_{p_{s}}(k_{s})\hat{A^{\dagger}}_{p_{i}}(k_{i}),
\end{equation}

where the coupling efficiency $\Gamma_{p_{s,i}}(k_{s},k_{i})$ to the Bloch wave results in the contribution of $N$ sublattices of the form:

\begin{equation}
 \Gamma_{p_{s,i}}(k_{s},k_{i}) = \gamma\sum_{n=1}^{N} E_{n}^{p}(k_{s}+k_{i})u_{p_{s},j}(k_{s})u_{p_{i},j}(k_{i}).
\end{equation}

For the case of a pump mode coupled to only two sublattices labeled by  $n=1,2$, substituting the eigen-mode profile determined by the topology of the Quantum Walk Hamiltonian $u_{s,i}$ (Eq. 8), we obtain an expression for $\Gamma_{p_{s,i}}(k_{s},k_{i})$ of the form:

\begin{equation}
 \Gamma_{p_{s,i}}(k_{s},k_{i}) = \gamma (E_{1}^{p}(k_{s}+k_{i})+ E_{2}^{p}(k_{s}+k_{i})\frac{n_{x}(k)+in_{y}(k)}{n_{z}(k)\pm\lambda(k)}).
\end{equation}

For the particular case that $n_{z}=0$ we obtain the simplified expression:

\begin{equation}
 \Gamma_{p_{s,i}}(k_{s},k_{i}) = \gamma (E_{1}^{p}(k_{s}+k_{i})e^{-i|\phi_{s}(k)+\phi_{i}(k)|}+ E_{2}^{p}(k_{s}+k_{i})),
\end{equation}

Where $\phi_{s,i}(k)$ are the phase matching functions for signal and idler, which depend on the quantum walk topology. For the particular case of the split-step quantum walk, this results in Eq. (9), for each of the bi-photons independently. 

\section{Numerical Results}

We first performed simulation in order to quantify the impact of the pump envelope $E_{1}^{p}(k_{s}+k_{i})$ on the coupling efficiency $\Gamma_{p_{s,i}}(k_{s},k_{i})$. This in turn can provide information about the spatial correlations between the bi-photons produced by SPDC, since the efficiency is proportional to the probability amplitude of emission of bi-photons. In particular, a tilted coupling efficiency parameter in the $k_{s,i}$-plane,  will characterize  spatial correlations or anti-correlation between the bi-photons. In all simulations we assumed a sufficiently large crystal length $L$, so that the $\mathrm{sinc}$ dependence of the phase-matching function is maximal and can be considered constant.   This is illustrated in Fig. 1, for ($\theta_{1}^{s,i}=0.01, \theta_{2}^{s,i}=0.0001$) and for two different pump envelope widths $\sigma=500$ and $\sigma=10$, and for different relative phases between the signal and idler $\phi_{s}(k)=\pm \phi_{i}(k)$. Fig. 1 (a) $\theta_{1}^{s,i}=0.01, \theta_{2}^{s,i}=0.001 $, $\phi_{s}(k)= \phi_{i}(k)$ and pump envelope width $\sigma=500$, Fig. 1 (b) $\theta_{1}^{s,i}=0.01, \theta_{2}^{s,i}=0.001 $, $\phi_{s}(k)=-\phi_{i}(k)$ and pump envelope width $\sigma=500$, Fig. 1 (c) $\theta_{1}^{s,i}=0.01, \theta_{2}^{s,i}=0.001 $, $\phi_{s}(k)= \phi_{i}(k)$ and pump envelope width $\sigma=10$, Fig. 1 (d)$\theta_{1}^{s,i}=0.01, \theta_{2}^{s,i}=0.001 $, $\phi_{s}(k)= -\phi_{i}(k)$ and pump envelope width $\sigma=10$. It is apparent that a small envelope reduces coupling efficiency to only the larges values of momentum for signal and idles. On the other hand, as expected, as the width $\sigma$ of the pump envelope increases so does the coupling efficiency.  \\

\begin{figure} [t!]
\label{fig:1}
\includegraphics[width=0.9\linewidth]{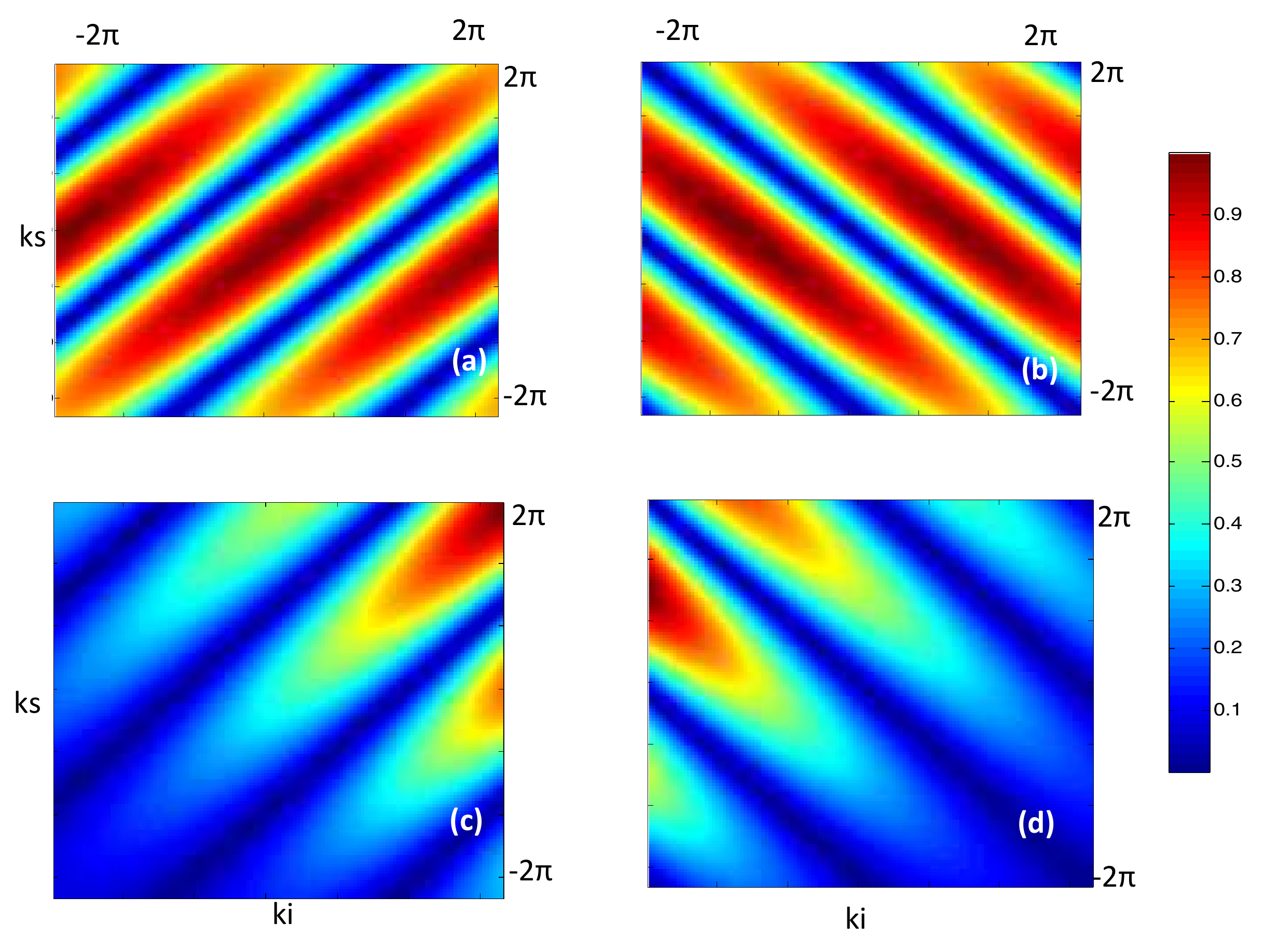} \caption{Numerical simulation of coupling efficiency $ \Gamma_{p_{s,i}}(k_{s},k_{i})$ in Fourier Domain. (a) $\theta_{1}^{s,i}=0.01, \theta_{2}^{s,i}=0.001 $, $\phi_{s}(k)= \phi_{i}(k)$ and pump envelope width $\sigma=10$, (b) $\theta_{1}^{s,i}=0.01, \theta_{2}^{s,i}=0.001 $, $\phi_{s}(k)=-\phi_{i}(k)$ and pump envelope width $\sigma=10$, (c) $\theta_{1}^{s,i}=0.01, \theta_{2}^{s,i}=0.001 $, $\phi_{s}(k)= \phi_{i}(k)$ and pump envelope width $\sigma=500$, (d)$\theta_{1}^{s,i}=0.01, \theta_{2}^{s,i}=0.001 $, $\phi_{s}(k)= -\phi_{i}(k)$ and pump envelope width $\sigma=500$.}
\end{figure}

In order to further analyze the impact of the quantum walk lattice topology on the type of coupling efficiency that can be expected, we performed simulations considering a constant amplitude for the pump $E^{p}(k_{s}+k_{i})=E^{p}$, with no dependence on $k$ in Fourier Space. We consider two cases, corresponding to phase parameters $\theta_{1,2}^{s,i}$ defining different phase matching conditions $\phi_{s,i}(k)$ for signal and idler photons. This is illustrated in Fig. 2: Fig. 2 (a) $\theta_{1}^{s,i}=0.01, \theta_{2}^{s,i}=9\times \pi/20 $ and $\phi_{s}(k)=  \phi_{i}(k)$, Fig. 2 (b) $\theta_{1}^{s,i}=0.01, \theta_{2}^{s,i}=0.001 $ and $\phi_{s}(k)= \phi_{i}(k)$, Fig. 2 (c) $\theta_{1}^{s,i}=0.01, \theta_{2}^{s,i}=9\times \pi/20 $ and $\phi_{s}(k)=  -\phi_{i}(k)$, Fig. 2 (d) $\theta_{1}^{s,i}=0.01, \theta_{2}^{s,i}=0.001 $ and $\phi_{s}(k)= -\phi_{i}(k)$. Fig. 2 reveals the emergence of a non-trivial 2D-imensional grid in the coupling efficiency for quantum walk lattice parameters in distinct topological sectors (Fig. 2 (a) and (c)). The periodicity in the grid is a clear consequence of the periodicity in lattice parameters and $k$-space. On the other hand for lattice parameters in the same topological sector (Fig. 2 (b) and (d)), we obtain the same type of coupling coefficient as expected for standard SPDC, of course with no $k$-dependence since this was ignored in the approximation of constant pump envelope $E^{p}(k_{s}+k_{i})=E^{p}$.

\begin{figure} [t!]
\label{fig:1}
\includegraphics[width=0.9\linewidth]{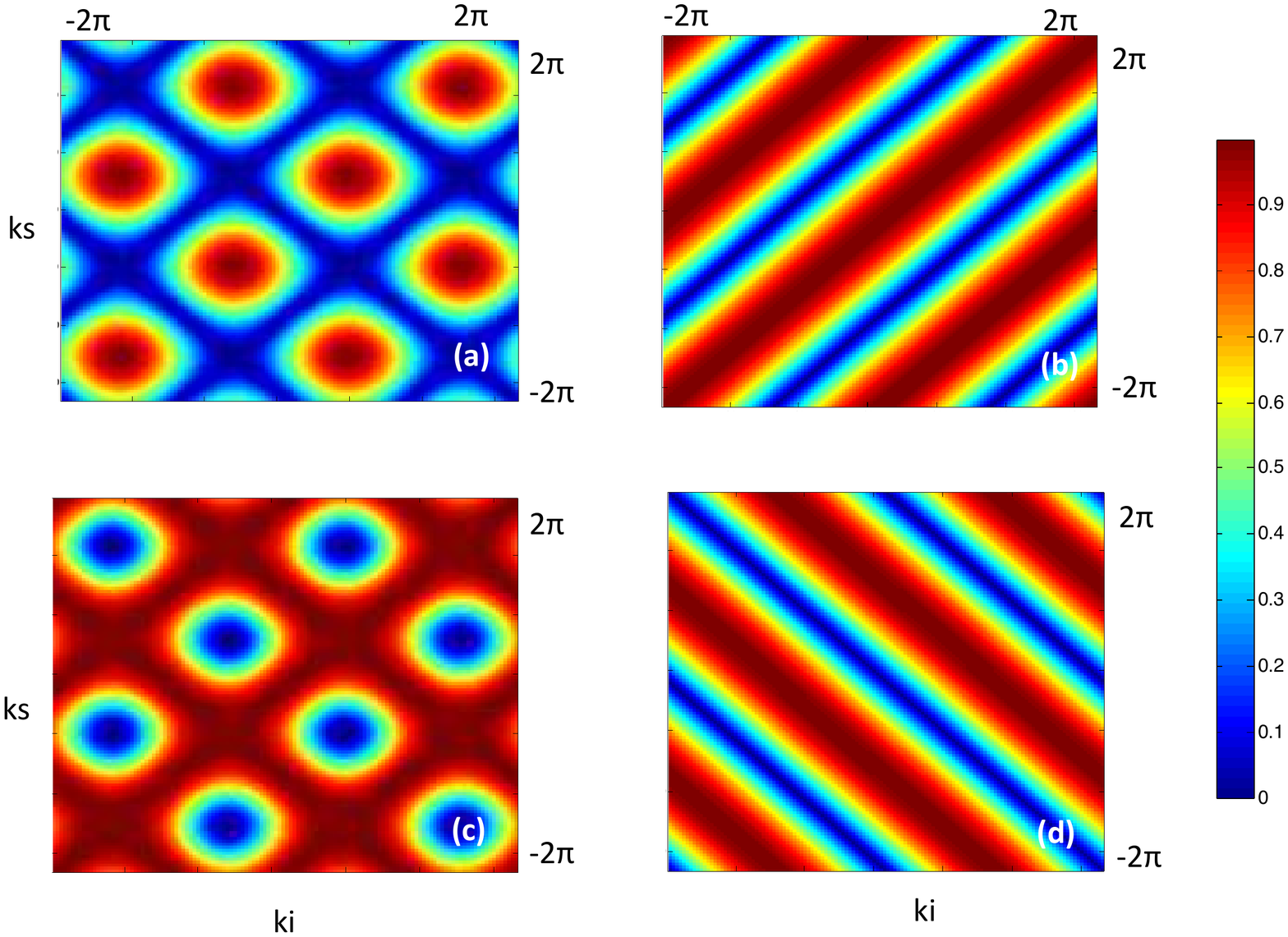} \caption{Numerical simulation of coupling efficiency $ \Gamma_{p_{s,i}}(k_{s},k_{i})$ in Fourier Domain. (a) $\theta_{1}^{s,i}=0.01, \theta_{2}^{s,i}=9\times \pi/20 $ and $\phi_{s}(k)=  \phi_{i}(k)$, (b) $\theta_{1}^{s,i}=0.01, \theta_{2}^{s,i}=0.001 $ and $\phi_{s}(k)= \phi_{i}(k)$, (c) $\theta_{1}^{s,i}=0.01, \theta_{2}^{s,i}=9\times \pi/20 $ and $\phi_{s}(k)=  -\phi_{i}(k)$, $\theta_{1}^{s,i}=0.01, (d) \theta_{2}^{s,i}=0.001 $ and $\phi_{s}(k)= -\phi_{i}(k)$.}
\end{figure}

\section{Experimental Methods}

\subsection{\emph{Fiber Network}}

\noindent In Ref \cite{photons2} the authors performed an optical implementation of the operator defined by Eq. (2), using polarization degrees of freedom of single photons and a sequence of half-wave plates and calcite beam-splitters. On the other hand, in Ref \cite{Peruzzo} the authors implemented a quantum walk in a lattice of coupled wave-guides. 
In this work we propose to use a fiber network to implement a quantum walk to simulate 1D and 2D  topological phases. One of the main ingredients is the implementation of an optical non-linearity which can introduce the production of bi-photons, for instance via the process of Spontaneous Parametric Down Conversion (SPDC) as described in the previous sections. We argue that in this way, one could simulate both attractive and repulsive interactions (for the case of correlated or anti-correlated down-converted bi-photons). A similar idea was already proposed in \cite{Silberberg}, where attractive interactions were introduced in a planar AlGaAs waveguide characterized by  a strong focussing Kerr non-linearity. Likewise, repulsive interactions can be simulated using defocusing non-linearities \cite{Silberberg}, though this would remain part of future efforts. A suitable alternative kind of waveguide for the simulation of attractive interactions are photonic band gap fibers with a Raman active gas, which are predicted to have a strong non-linearity. These fibers consist of a hollow core photonic crystal fiber filled with an active Raman gas which are capable of exceeding intrinsic Kerr non-linearities by orders of magnitude   \cite{Skryabin, Peschel}.

\subsection{\emph{Input state preparation}}

\noindent For the linear (non-interacting) case (QW with SU(2) symmetry) we plan to use single-mode states both with Poissonian and Subpoissonian statistics, such as coherent states and squeezed coherent  or single-photon states. The non-classical nature of the squeezed and single-photon states should be revealed in the intensity distribution of counts as well as in the standard deviation. On the other hand, for the non-linear (interacting) case (SPDC with SU(1,1) symmetry) quantum theory  predicts that the probability amplitudes of the modes should interfere leading to an enhancement/reduction of the initial correlations. One of the goals of the project is to analyze the sensitivity of the non-linear network to phase relations dictated by the topology of the network in the input state and to the amount of gain. We also plan to analyze the effect of correlations and entanglement in the input state on localized edge states and to find some kind of non-local order parameter characterizing topological order \cite{Haldane}. Finally, one of the aims of this research plan is to demonstrate the feasibility of entanglement topological protection \cite{PuentesJPB2012}.\\

\subsection{\emph{Detection Schemes}}

\begin{itemize}
\item{\bf Intensity probability distributions and standard deviation}\\
\noindent The most direct form of measurement is to detect the statistics of counts by studying intensity histograms of photons and their standard deviation, as described in Ref \cite{photons2, Silberberg}. In particular, by placing a photo-diode/APD at the output of each fiber,  characterizing a given site in the network, it is possible to obtain a probability distribution of counts and its standard deviation along the $N$ steps of the quantum walk.  While in the case of input states with Poissonian statistics we expect to find a classical \emph{binary} distribution of counts as the output of the quantum walk, in the non-classical case we expect to find a localized edge state at the boundary between two different topological sectors. Furthermore, we plan to measure the normalized standard deviation $\sigma_{N}$ for the classical and non-classical case, where we expect to find a markedly different dependance on the number of steps $N$; namely, while for the classical walk (coherent states)  we plan to obtain   $\sigma_{N} \propto \sqrt{N}$ diffusive dependance, for the non-classical case  (squeezed states, single photons) we plan to obtain an $\sigma_{N} \propto N $ ballistic dependance with the number of discrete steps.

\item{\bf HBT correlation measurements}\\
\noindent When using non-linear fibers in the amplifying network, it would be interesting to analyze 1-mode ($g^{1}(\Delta r)$) and 2-mode ($g^{2}(\Delta r)$) spatial correlations functions by means of Hanbury-Brown-Twiss (HBT) like interferometers between different output modes in the network, as described in Ref \cite{Silberberg}.  In particular, while in the case of attractive interactions, as simulated by Kerr non-linearities in fibers, the correlations are expected to increase, for the repulsive case the correlations are expected to decrease. We also plan to analyze the dependance of spatial correlations  on the amount of gain present in the medium. In particular, for some critical value of the overall gain $G_{c}$ we expect to find a decay of the correlations, which in turn can be related to the classical-quantum transition in amplifying media \cite{Torma,Peschel}. Finally, we will also investigate the response of the amplifying network to different phases in the input states dictated in turn by the topology of the network, as well as the response to phase noise, by introducing phase averaging mechanisms (see Fig. 3).

 \end{itemize}
 
\begin{figure} [t!]
\label{fig:1}
\includegraphics[width=0.9\linewidth]{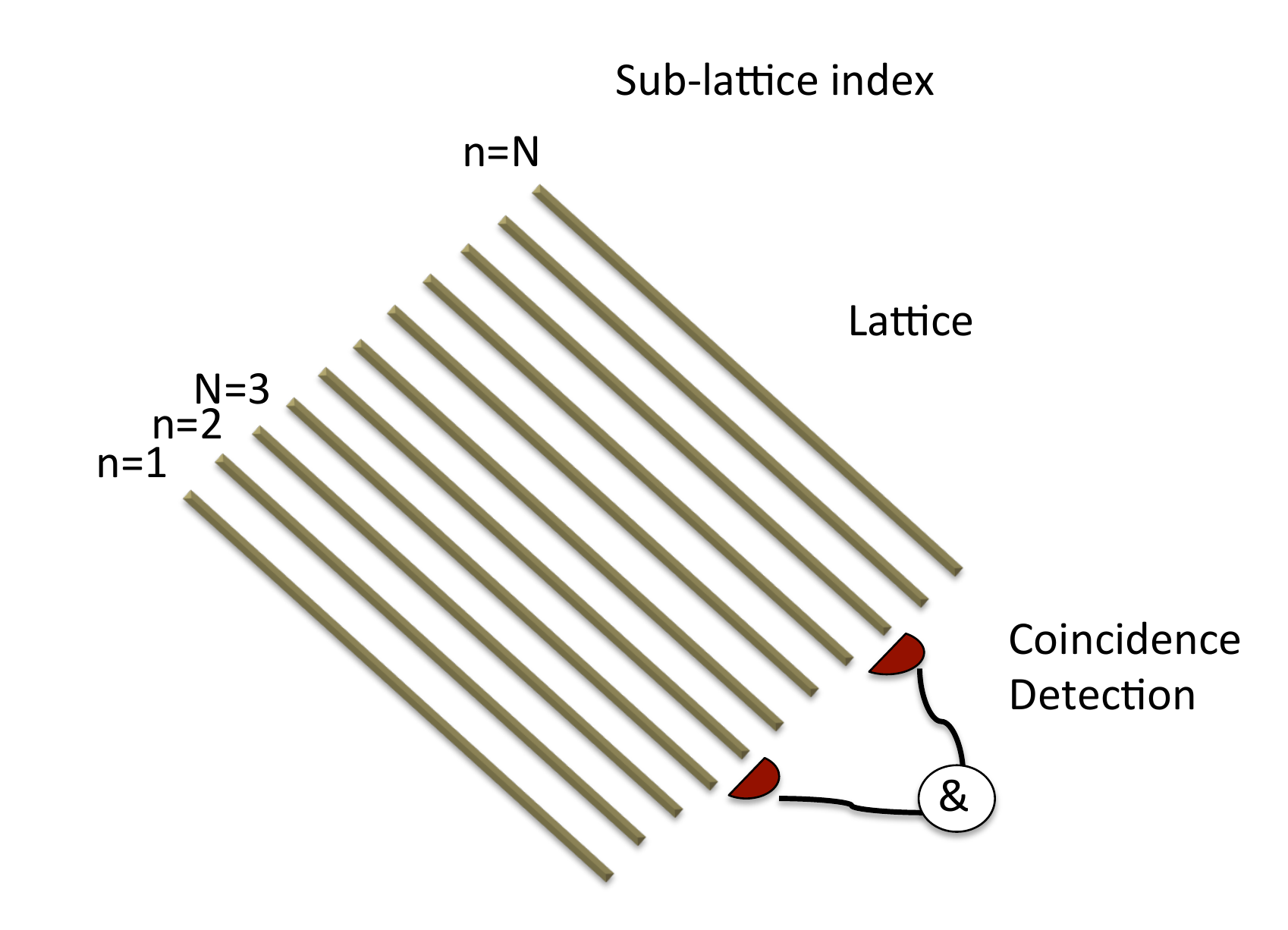} \caption{Experimental setup for measurement of spatial correlation via coincidence counts between different sublattice modes ($n$). }
\end{figure}

\section{DISCUSSION AND OUTLOOK}

\noindent In this work we propose an experimental implementation of topological phases by means of an optical implementation of a discrete time quantum walk architecture (DTQWs). One of the main novel ingredients is the inclusion of non-linear media and non-linear effects in the DTQW via the possibility of spontaneous parametric down conversion (SPDC) in the lattice. By means of numerical simulations, we have analyzed the interplay between quantum walk topology and spatial properties of photon pairs produced by spontanenous parametric down conversion. In particular, have numerically described how the topology of the quantum walk lattice can play an important role in the phase-matching function of bi-photons produced by spontaneous parametric down conversion. As a future work,  we expect to characterize the robustness of such topological phases and their characteristic bound states against amplitude and phase noise as well as to decoherence, by tracing over spatial modes of the field. One of the main goals of the proposed work is to investigate the use of parametric amplifiers as a means of simulating many-body effects in topological phases. In particular, we expect to link such phases with the classical or quantum statistics of the fields by means of intensity distribution and spatial correlation measurements, and we intend to find a link between some measure of entanglement and a non-local order parameter characterizing the topology of the phase \cite{Haldane}, or the feasibility of entanglement topological protection approaches \cite{PuentesJPB2012}. Some significant signatures of many-body dynamics in topological order are expected to be apparent, such as charge fractionalization and Hall quantization, which motivate the extension of the research to the non-linear (many-body) scenario. Furthermore, other more complex topological phases (such as spin Hall phase) could  be simulated in the future by all optical means by using 2D quantum walks and higher dimensional internal degrees of freedom of the radiation field, such as  the orbital angular momentum \cite{PuentesPRL2012}. Furthermore, topological order has been  considered as a useful ingredient  for fault tolerant quantum computation, as it can protect the system against local perturbations which would otherwise introduce decoherence and loss of quantum information  \cite{Preskill}. 


\end{document}